\documentclass[onecolumn,fleqn,10pt]{wlscirep}
\usepackage{dcolumn}
\usepackage{bm}
\usepackage{epsfig}
\usepackage{graphicx}
\usepackage{epstopdf}
\usepackage{amssymb}
\usepackage{amsfonts}
\usepackage{hyperref}
\usepackage{color,soul}
\usepackage{appendix}
\usepackage{wrapfig}

\title{Fast optical cooling of a nanomechanical cantilever by a dynamical
Stark-shift gate}

\author[1]{Leilei Yan}
\author[2,*]{Jian-Qi Zhang}
\author[3]{Shuo Zhang}
\author[1,$\dagger$]{Mang Feng}
\affil[1]{State Key Laboratory of Magnetic Resonance and Atomic and Molecular Physics, Wuhan Institute of Physics and Mathematics, Chinese Academy of Sciences, Wuhan, 430071, China}
\affil[2]{University of the Chinese Academy of Sciences, Beijing 100049, China}
\affil[3]{Zhengzhou Information Science and Technology Institute, Zhengzhou, 450004, China}

\affil[*]{changjianqi@gmail.com}
\affil[$\dagger$]{mangfeng@wipm.ac.cn}

\begin{abstract}
The efficient cooling of the nanomechanical resonators is essential to exploration of quantum properties of the macroscopic or mesoscopic systems. We propose such a laser-cooling scheme for a nanomechanical cantilever, which works even for the low-frequency mechanical mode and under weak cooling lasers. The cantilever is coupled by a diamond nitrogen-vacancy center under a strong magnetic field gradient and the cooling is assisted by a dynamical Stark-shift gate. Our scheme can effectively enhance the desired cooling efficiency by avoiding the off-resonant and undesired carrier transitions, and thereby cool the cantilever down to the vicinity of the vibrational ground state in a fast fashion.
\end{abstract}
\begin{document}

\flushbottom
\maketitle
%
%



Over the past years, nano-mechanical resonators (NRs) have attracted
considerable attention both theoretically and experimentally and presented
potential applications based on the quantum properties, for example,
optomechanical induced transparency \cite{Sicience-330-1520}, photon
blockade \cite{nature-436-87,prl-107-063601}, optical Kerr effect \cite%
{pra-32-2287}, entanglement between microscopic objects \cite%
{V.B.Braginsky,prb-88-085201}, quantum state measurement \cite%
{prb-68-155311,nature-459-960}, biological sensing detection \cite%
{natureN-3-691,natN-3-501} and hybrid coupling to cold atoms \cite%
{prl-99-140403}.

However, the quantum properties regarding NRs are always hidden by
the thermal phonons involved. To suppress the thermal phonons, many
schemes have been proposed so far to try to cool the NRs down to the
vicinity of their vibrational ground states, such as the sideband
cooling \cite {prl-92-075507,nature-475-359}, the backaction
sideband cooling \cite
{prb-78-134301,nature-432-200,prl-99-093901,prl-99-093902,prb-76-205302,nature-443-193}
, the hot-thermal-light-assisted cooling \cite{prl-108-120602}, the
time-dependent control cooling \cite{JPCM-25-142201,pra-83-043804},
the quadratic-coupling-based cooling \cite{pra-85-025804}, the
measurement cooling \cite{prb-84-094502} and the electromagnetically
induced transparency (EIT) cooling \cite
{epl-95-40003,prl-103-227203,pra-85-033835,pra-86-053828,oe-21-029695}.

The EIT cooling works based on quantum interference, which enhances
the first-order red-sideband transition for cooling, but eliminates the carrier
transition and suppresses the first-order blue-sideband transition for heating
\cite
{epl-95-40003,prl-103-227203,pra-85-033835,pra-86-053828,oe-21-029695}.
In particular, it works efficiently even in the non-resolved
sideband regime in the laboratory representation, i.e., with the large spontaneous emission rate. The
EIT cooling was first proposed and experimentally implemented in the
trapped-ion system \cite {prl-85-4458,prl-85-5547}, and then
extended to other systems, including the quantum dot
\cite{epl-95-40003,pra-85-033835,pra-86-053828}, the superconducting
flux qubit \cite{prl-103-227203} and the diamond nitrogen-vacancy
(NV) center \cite{oe-21-029695}. However, for the Rabi frequency
comparable to the vibrational frequency of the NR, the prerequisite of the fast
cooling, the existing cooling scheme could not work efficiently
\cite {oe-21-029695}. Therefore, developing an alternative scheme
available for the NR cooling, which is faster and more efficient
than the EIT cooling, is highly desirable \cite{njp-9-279}. On the
other hand, a NV center coupled to a nanomechanical cantilever can
be used to cool the cantilever vibration down to a quantum regime
\cite{oe-21-029695,prb-79-041302}, where the coupling is from the
magnetic field gradient (MFG). The extension of such a coupling is
applicable to future scalable quantum information processor \cite
{prb-79-041302,natphys-7-879}. To this end, achievement of high-quality NRs and
cooling low-frequency NRs are highly expected, but experimentally demanding.

The present work focuses on the ground-state cooling of a NR with the assistance from a Stark-shift
gate in the non-resolved sideband regime in the laboratory representation. Such a cooling scheme can cool a low-frequency ($\leq$1 MHz)  NR more efficiently than the
conventional sideband cooling due to elimination of the involved
carrier transitions which contribute for heating, as confirmed for cooling the trapped ion \cite{njp-9-279}. However, compared to
the trapped ion, the NR (i.e., the cantilever) under our
consideration is of a much bigger mass, which decreases the
NV-cantilever coupling to be nearly zero. To generate a strong
enough coupling between the NV center and the cantilever, we
introduce a strong MFG \cite{oe-21-029695,prb-79-041302}.
Moreover, since the cantilever is more sensitive to the
environmental noise than the trapped ion, we have to seriously
consider the influence from the non-zero temperature thermal noise
of the environment in our calculation.

The key point in the present work is the introduction of an effective
classical filed  to couple the sublevels of the electronic ground state of the NV
center, which creates a dynamical Stark shift under the strong MFG and
accelerates the cooling of the cantilever by suppressing the undesired
transitions. We show the possibility to cool the cantilever with the same
cooling rate as in the trapped-ion system \cite{njp-9-279}. Moreover,
different from the microwave cooling scheme \cite{prb-79-041302}, in
which the magnetic tip with the fixed MFG is attached at the end of the
cantilever, the cooling in our case is made by lasers, which ensure that the cooling rate (the cooling speed)
in our scheme is larger (faster) than the one based on the microwave\cite{prb-79-041302}, and the MFG in our idea is
independent from the cantilever, but
generated by the coils and controlled by the external electric current. The MFG
in our design is evidently more flexibly adjustable.

More specifically, we show below that the addition of the Stark-shift gate makes the
cooling more powerful than the optics-based EIT cooling in a
previous scheme \cite {oe-21-029695}, and is particularly useful for
the cantilever of lower vibrational frequency under weaker laser
irradiation. Since the cooling of the low-frequency cantilevers down
to the ground state is still challenging with current technology, and
the requirement of weak laser irradiation can reduce the
experimental difficulty, our scheme is of practical application in
exploring quantum properties of the nanomechanical cantilevers.


\section*{Results}

\begin{figure}[tbp]
\centering
\includegraphics[width=14cm,height=6cm]{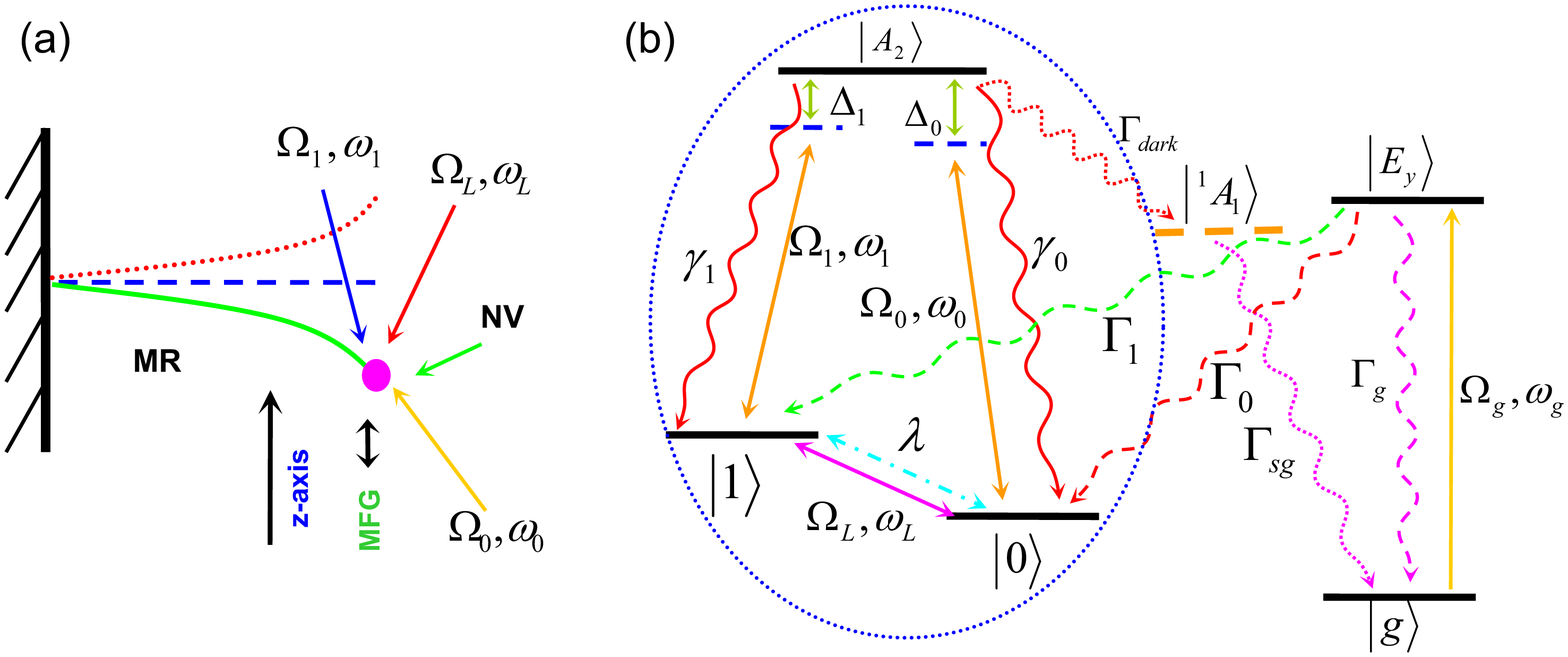}
\caption{(a) Schematic illustration of our cooling scheme using a dynamic
Stark-shift gate, where the nanomechanical cantilever is attached by a NV
center under irradiation of lasers and an effective classical field. (b) The encircled
three levels form our major part of the cooling, where the irradiation of
the two lasers satisfies the two-photon resonance with $\Delta_{0}=
\Delta_{1} $ and the effective classical field is in resonance with the
transition between the two ground states, i.e., $\protect\omega_L=\protect%
\omega_0-\protect\omega_1$. The cantilever vibration is coupled to the NV
center by a strong MFG. $\protect\gamma_{0}$ ($\protect\gamma_{1}$) is the
decay from the excited state $|A_2\rangle$ to the ground state $|0\rangle$ ($%
|1\rangle$). The auxiliary part is outside the circle, which includes a
ground state $|g\rangle$, an excited state $|E_y\rangle$ and the metastable
state $|^1A_1\rangle$. There is a pumping field with the frequency $\protect%
\omega_g$ to couple $|g\rangle$ with $|E_y\rangle$ by the Rabi frequency $%
\Omega_g$. $\Gamma_{dark}$ is the decay rate from $|A_2\rangle$ to $%
|^1A_1\rangle$, and $\Gamma_0$ ($\Gamma_1$, $\Gamma_g$) is the decay rate
from the excited state $|E_y\rangle$ to the ground state $|0\rangle$ ($%
|1\rangle$, $|g\rangle$), and $\Gamma_{sg}$ represents the decay from $%
|^1A_1\rangle$ to $|g\rangle$ (see Appendix in the EIT cooling scheme \cite {oe-21-029695} for details).}
\label{NVcooling}
\end{figure}

\noindent\textbf{The cooling of a NV-cantilever system by a Stark-shift gate.}
The model of our system is presented in figure \ref{NVcooling}(a), where a negatively charged NV center is attached at the end of a nanomechanical
cantilever under a strong MFG. The ground state of the NV center is a spin
triplet with a zero-field splitting $2\pi\times 2.87$ GHz between $m_{s}=0$
and $m_{s}=\pm 1$, where $m_{s}$ is the projection of the total electron
spin $S=1$ along the $z$-axis. The sublevels $m_{s}=\pm 1$ are employed for
qubit encoding in our cooling scheme, with $m_{s}=-1$ as $|0\rangle$ and
$m_{s}=+1$ as $|1\rangle$. $m_{s}=0$ is labeled as an auxiliary level
$|g\rangle$ for preventing leakage from the excited state $|A_{2}\rangle$, as
discussed below. According to the selection rule of the transitions \cite%
{njp-13-025025,nature-466-730}, the state $|0\rangle$ ($|1\rangle$) may be
coupled to the excited state $|A_{2}\rangle$ by a polarized laser \cite {oe-21-029695,njp-13-025025,
nature-466-730, pra-83-054306}. $|A_{2}\rangle$ is an entangled state
involving the components $|0\rangle $, $|1\rangle$ and the orbital states,
and keeps separate enough from neighboring levels even under the strain \cite{nature-466-730}.
The state $|0\rangle$ can be coupled to $|1\rangle$ by an effective classical field
due to two-photon Raman process (Adopted in the present work; see Appendix B for details) or
by a stress applied perpendicularly to the axial direction of the NV center \cite{prl-111-227602}.
In figure \ref{NVcooling}(b), the $\sigma_{-}^{0}$ ($\sigma _{+}^{1}$) polarized laser owns the frequency $%
\omega _{0}$ ($\omega _{1}$) and the Rabi frequency $\Omega _{0}$ ($\Omega
_{1}$), and the effective classical field is with the frequency $\omega _{L}$ and the Rabi
frequency $\Omega _{L}$.

Moreover, there are leakages from the excited state $|A_{2}\rangle$ down to
the metastable state $|^{1}A_{1}\rangle$, which would stop the cooling
process. To solve this problem, we employ $|g\rangle$ and another excited
state $|E_{y}\rangle$ as the auxiliary part in figure \ref{NVcooling}(b),
whose function can be found in the EIT cooling scheme \cite {oe-21-029695} and would not be
reiterated in the present paper. Furthermore, different from the
trapped-ion system, in which the coupling between the internal and the
vibrational degrees of freedom is caused by the mechanical effect of light
\cite{jpb-36-1041}, our model uses the MFG to provide the coupling between
the NV center and the vibration of the cantilever. The MFG consists of a
coil wrapping a permanent magnet core, controlled by the external electric
current.

The Hamiltonian of the system in units of $\hbar =1$ is given by
\begin{equation}
\begin{array}{ccl}
H & = & \omega _{k}a^{\dagger }a+\omega _{A}\left\vert A_{2}\right\rangle
\left\langle A_{2}\right\vert +g_{e}\mu _{B}B(0)(\left\vert 1\right\rangle
\left\langle 1\right\vert -\left\vert 0\right\rangle \left\langle
0\right\vert ) \\
& &+ \frac{\Omega }{2}(\left\vert A_{2}\right\rangle \left\langle
1\right\vert e^{-i\omega _{1}t}+\left\vert A_{2}\right\rangle \left\langle
0\right\vert e^{-i\omega _{0}t}+h.c.)+\frac{\Omega _{L}}{2}(\left\vert
0\right\rangle \left\langle 1\right\vert e^{i\omega _{L}t}+h.c.) \\
& &+ \lambda (\left\vert 1\right\rangle \left\langle 1\right\vert
-\left\vert 0\right\rangle \left\langle 0\right\vert )(a^{\dagger }+a),%
\end{array}
\label{C0}
\end{equation}
where $a^{\dagger}$ ($a$) is the creation (annihilation) operator of the
cantilever vibration with the frequency $\omega_{k}$, $\omega_{A}$ is
regarding $|A_{2}\rangle $, $B(0)$ is the constant magnetic field strength,
$g_{e}$ is the $g$-factor and $\mu_{B}$ is the Bohr magneton. $\Omega(=\Omega_{0}=\Omega_{1})$
and $\Omega_{L}$ are the Rabi frequencies regarding irradiation from the laser
and the effective classical field. $\lambda =g_{e}\mu_{B}B^{\prime}(0)x_{0}$ is the coupling
due to the MFG $B^{\prime}(0)$ with $x_{0}=1/\sqrt{2M\omega _{k}}$ and a
cantilever mass $M$ \cite{prl-87-257904,prb-79-041302,natphys-7-879}. Since the NV center
is sensitive to the strain, we suppose a low strain condition throughout this work, which
ensures that the employed excited state $|A_2\rangle$ robustly owns stable symmetry properties \cite{nature-466-730}.

To understand the cooling physical picture and simplify the calculation, we
make a unitary transformation on equation (\ref{C0}). In the rotating frame, we have $|\psi ^{\mathrm{rot}}(t)\rangle
=e^{-iRt}|\psi (t)\rangle $ and $H^{\mathrm{\ rot}}=e^{-iRt}H(t)e^{iRt}+R$
with $R\equiv \omega _{1}\left\vert 1\right\rangle \left\langle 1\right\vert
+\omega _{0}\left\vert 0\right\rangle \left\langle 0\right\vert $ \cite%
{Gardiner}. Under the near-resonance condition, equation (\ref{C0}) can be
rewritten in a time-independent form as $H^{\mathrm{rot}}=H_{0}+V$, with
\begin{equation}
H_{0}=\omega _{k}a^{\dagger }a-\Delta \left\vert A_{2}\right\rangle
\left\langle A_{2}\right\vert +\frac{\sqrt{2}}{2}\Omega (\left\vert
A_{2}\right\rangle \left\langle b\right\vert +h.c.)+\frac{1}{2}\Omega
_{L}(\left\vert b\right\rangle \left\langle b\right\vert -\left\vert
d\right\rangle \left\langle d\right\vert ),%
\label{C1}
\end{equation}
and
\begin{equation}
V=\lambda (\left\vert b\right\rangle \left\langle d\right\vert +\left\vert
d\right\rangle \left\langle b\right\vert )(a^{\dagger }+a),  \label{C2}
\end{equation}
where $\left\vert b\right\rangle =\frac{1}{\sqrt{2}}(\left\vert
1\right\rangle +\left\vert 0\right\rangle )$ and $\left\vert d\right\rangle =%
\frac{1}{\sqrt{2}}(\left\vert 1\right\rangle -\left\vert 0\right\rangle )$
are the corresponding bright and dark states, respectively, and the detunings
satisfy $\Delta\equiv \Delta _{0}=\Delta _{1}$ with $\Delta _{0}=\omega
_{0}-\omega _{A}-g_{e}\mu _{B}B(0)$ and $\Delta _{1}=\omega _{1}-\omega
_{A}+g_{e}\mu _{B}B(0)$. Moreover, the last term in equation (\ref{C1})
describes the energy difference between the bright and dark states caused by
the effective classical field for the Stark shift, by which a Stark-shift gate
will be performed below for the cooling of the cantilever vibration (see Methods). Besides,
the coupling between the cantilever and the NV center is created by the strong MFG, which makes
the first-order red-sideband transition dominate the cooling process, as shown below.
With the assistance of the effective classical field, the energy difference between the dark and bright states
is equal to the frequency of the cantilever vibration. As a result, the phonon
is dissipated by the coupling due to the MFG with the assistance of external fields.

\noindent\textbf{The analytical and numerical treatments for the cooling.} By utilizing the perturbation theory and the non-equilibrium
fluctuation-dissipation relation, the Hamiltonian $H^{rot}$ (equation (\ref{C1}) plus equation (\ref{C2}))
yields the heating (cooling) coefficient $A_{+}$ ($A_{-}$) caused by the
external fields as below,
\begin{equation}
A_{\pm }=\frac{2\Gamma \lambda ^{2}\Omega ^{2}}{[\Omega ^{2}+(\mp \omega
_{k}-\Omega _{L})(\pm 2\omega _{k}-2\Delta +\Omega _{L})]^{2}+\Gamma
^{2}(\mp \omega _{k}-\Omega_{L})^{2}},  \label{C21}
\end{equation}
whose deduction in details can be found in Supplementary Information. $\Gamma$ is the total decay rate regarding $|A_{2}\rangle$. The heating (cooling) coefficient
in equation (\ref{C21}) is different from the one obtained in other cooling methods\cite%
{prl-103-227203,oe-21-029695,prl-85-4458,pra-67-033402}, but can be reduced
to the result as the proposal for ion cooling \cite{njp-9-279} when $\Omega_{L}=\omega_{k}$. This
is due to the fact that both ion cooling scheme \cite{njp-9-279} and our scheme share
the same work point for the Stark-shift gate, related to the Rabi
frequency of the effective classical field  and the vibrational frequency
of the ion or cantilever. Nevertheless, the
cooling (heating) efficiency in the cooling scheme for ion \cite{njp-9-279} only depends on the Rabi
frequency of the microwave field, but ours is not only relevant to the effective
classical field Rabi frequency but also subject to the MGF coupling $\lambda$.

\begin{figure}[tbph]
\centering\includegraphics[width=8cm,height=4.5cm]{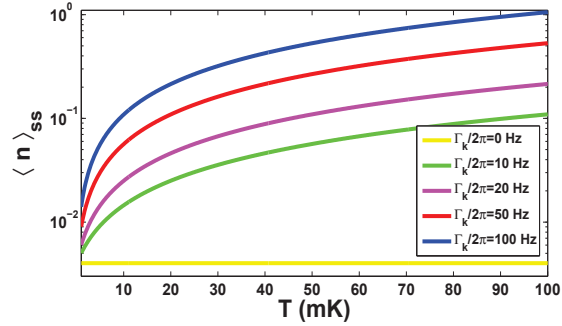} \centering
\caption{The final phonon number $\left\langle n\right\rangle _{ss}$ (in logarithmic scale) versus
the environmental temperature $T$ and the vibrational decay rate $\Gamma_k$,
where we have taken the parameter values $\protect\omega_{k}/2\protect\pi=2$
MHz, $\Omega/2\protect\pi=2$ MHz, $\Omega_L=\protect\omega_k$ and $\Delta =0$
\protect\cite{njp-13-025025}, as well as $\Gamma/2\protect\pi=15$ MHz and $%
\protect\lambda/2\protect\pi =0.115$ MHz \protect\cite%
{prb-79-041302,pra-72-043823}.}
\label{simu}
\end{figure}

A proper analysis of the phonon dissipation must consider the non-zero
temperature environmental noise, since the cantilever with much larger
volume and mass is more sensitive to the environment than the trapped ion.
Using the methods developed previously \cite{prl-103-227203,oe-21-029695}, we obtain
following analytical expression of the time-dependent average phonon number,
\begin{equation}
\left\langle n(t)\right\rangle =\left\langle n\right\rangle
_{ss}+e^{-(W+\Gamma _{k})t}[\left\langle n(0)\right\rangle -\left\langle
n\right\rangle _{ss}],  \label{C90}
\end{equation}
where the cooling rate $W=A_{-}-A_{+}$ originates from the interaction between the NV center and the
external fields \cite{prl-85-4458,pra-67-033402}. The final average
phonon number is
\begin{equation}
\begin{array}{l}
\left\langle n\right\rangle _{ss}=[A_{+}+K(\omega _{k})\Gamma
_{k}]/(W+\Gamma _{k}),%
\end{array}
\label{C10}
\end{equation}
where $\Gamma _{k}\equiv\omega _{k}/q$ is the vibrational decay rate with the
quality factor $q$ of the cantilever, and $K(\omega _{k})=1/(e^{\hbar \omega
_{k}/k_{B}T}-1)$ is the thermal occupation for the cantilever vibrational
degrees of freedom \cite{prl-103-227203} with the Boltzmann constant $k_{B}$
and the environmental temperature $T$, respectively. Figure \ref{simu} clearly presents
a better cooling effect occurring at the smaller decay rate $\Gamma_{k}$
and the lower environmental temperature $T$ compared with the previous scheme \cite{oe-21-029695}.

\begin{figure}[tbph]
\centering\includegraphics[width=12cm,height=5cm]{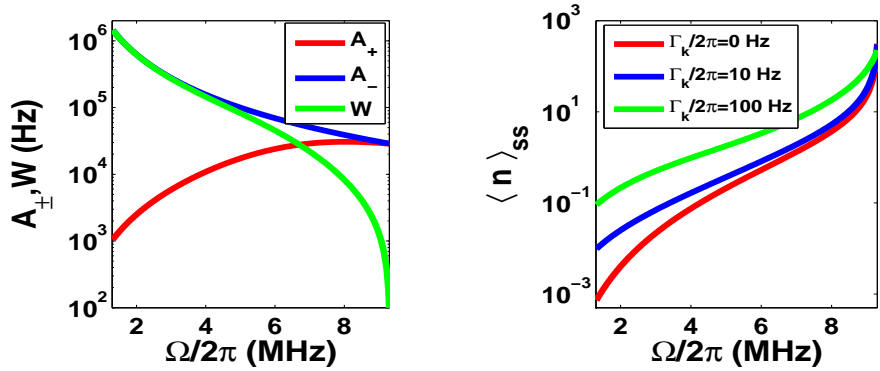} \centering
\caption{(Left) The heating coefficient $A_{+}$, the cooling coefficient $A_{-}$, and the
cooling rate $W=A_{-}-A_{+}$ (in logarithmic scale) versus $\Omega/2\pi$. (Right) The final
average phonon number $\left\langle n\right\rangle_{ss}$ (in logarithmic scale) versus
$\Omega/2\pi$ for different $\Gamma_k$, where we consider $\Omega_L=\protect%
\omega_k$ and the other parameters take the values as $\protect\omega_{k}/2%
\protect\pi =2$ MHz, $\Gamma/2\protect\pi=15$ MHz, $T=20$ mK, $\protect%
\lambda/2\protect\pi=0.115$ MHz and $\Delta=0$.}
\label{moW}
\end{figure}

For a deeper understanding of our cooling scheme, we may focus on the work
point $\Omega _{L}=\omega _{k}$ of the Stark-shift gate, which simplifies
the heating and cooling coefficients in equation (\ref{C21}) to be
\begin{equation}
A_{+}=\frac{2\Gamma \lambda ^{2}\Omega ^{2}}{4\Gamma ^{2}\omega
_{k}^{2}+(\Omega ^{2}-6\omega _{k}^{2}+4\Delta \omega _{k})^{2}}%
,~~~~~~~~~A_{-}= \frac{2\Gamma \lambda ^{2}}{\Omega ^{2}}.  \label{C11-2}
\end{equation}
As plotted in figure \ref{moW}(Left), $A_{+}$ ($A_{-}$) increases (decreases)
with $\Omega$. To make sure an efficient cooling, we should
have $A_{-}$ to be larger than $A_{+}$, implying an upper limit $\Omega
^{2}\leq M_{2}=\omega _{k}^{2}(\Gamma ^{2}+4\Delta ^{2}+9\omega
_{k}^{2}-12\Delta \omega _{k})/(3\omega _{k}^{2}-2\Delta \omega _{k})$ from
the above analytical expressions. Moreover, both figures \ref{moW}(Left) and
\ref{moW}(Right) show that the faster cooling and the minimal final phonon
number prefer smaller laser Rabi frequency. The extreme case happens at $%
\Omega =0$, in which we have $W=A_{-}$ due to negligible $A_{+}$, and
thereby $\left\langle n\right\rangle _{ss}$ tends to minimum. However, this
is a non-physical condition since $\Omega =0$ means no laser irradiation. In
our case, if the cooling works, $\Omega ^{2}>M_{1}=\max {\ [\Gamma\lambda, \omega _{k}\lambda,
\Delta\lambda ]}$ (meaning the internal dynamics faster than the external dynamics, e.g., $\sqrt{M_{1}} /2\pi =1.3$ MHz
in figure \ref{moW}) should be satisfied. Therefore, we reach a trade-off
regime for the laser irradiation $M_{1}<\Omega ^{2}\leq M_{2}$. On the other
hand, the laser detuning $\Delta$ involved in $A_{+}$ also has influence on
the cooling. To have a larger cooling rate, the larger blue detuning (i.e., $%
\Delta>0$) is required for the condition $\Omega ^{2}-6\omega_{k}^{2}>0$,
while the larger red detuning ($\Delta <0$) is necessary when $\Omega
^{2}-6\omega _{k}^{2}\leq 0$ is satisfied.

\begin{figure}[tbph]
\centering
\includegraphics[width=14 cm,height=7cm]{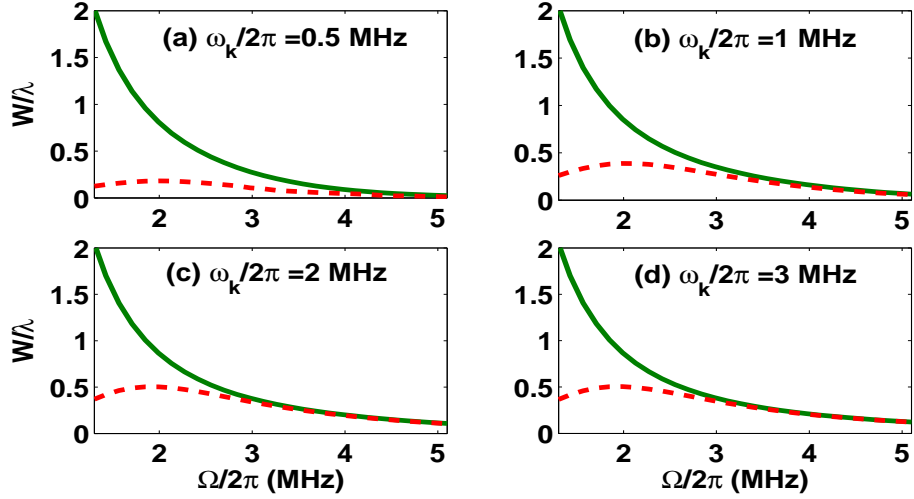}
\caption{ The cooling rate $W/\protect\lambda$ versus the Rabi frequency $\Omega/2\protect\pi$
where the work point $\Omega_L=\protect\omega_k$ is satisfied. The parameters take values as
$\protect\lambda/2\protect\pi=0.115 $ MHz, $\Gamma/2\pi=15$ MHz and $\Delta=0$. The solid
curves are the analytical results of the cooling rate $W$ by
equation (\protect\ref{C11-2}) and the dashed curves are
plotted by solving the master equations of $H^{rot}$. } \label{steady}
\end{figure}

The analytical results above (i.e., equations (\ref{C21}) and (\ref{C11-2})) are obtained under the perturbation and the
adiabatic condition. This implies that the real cooling effect should be
justified by the small values of $\Omega$, for which the adiabatic
condition is not fully satisfied. To this end, we have numerically
calculated the cooling rate $W$ with respect to $\Omega$ at the work point
of the Stark-shift gate. We may find from the figure \ref{steady} that
the discrepancy between the analytical and numerical results appears
within the regime $\Omega /2\pi <3$ MHz where the adiabatic condition is no
longer valid. If we check this regime more carefully, we find that the
discrepancy is bigger for the lower frequency of the cantilever.

The physical reason for the cooling rates plotted in figure \ref{steady} can be
understood by the decay and the pumping in the cooling process. Since
the transition between the bright and dark states is dipolar forbidden, we
excite the system from the bright state $|b\rangle$ to the excited state $|A_{2}\rangle$%
, and then decays down to the dark state $|d\rangle$. With a stronger pumping light,
the effective decay from the bright state to the dark state would be bigger, which yields a larger cooling rate.
However, a much stronger light would shift the bright state and thereby
weaken the red-sideband transition. As a result, with increasing $\Omega$,
the cooling rate increases at first, and then decreases, as examined by
numerics in figure \ref{steady}. However, the analytical solutions in equation (\ref{C11-2})
could not exactly describe the above cooling process if $\Omega /2\pi <3$
MHz.

\begin{figure}[tbph]
\centering
\includegraphics[width=14 cm,height=7cm]{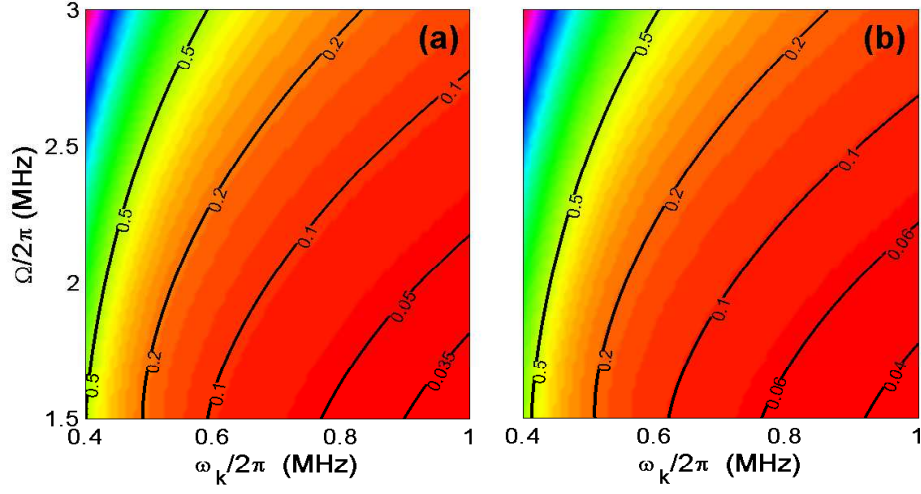}
\caption{ The final average phonon number $ \langle n\rangle_{ss} $ versus the vibrational frequency $\omega_{k}/2\pi$ and
Rabi frequency $\Omega/2\pi$, where the work point $\Omega_L=\protect\omega_k$ is satisfied. (a) The case of zero temperature
of environment without considering the vibrational decay. (b) The case of the environmental temperature $T=20$ mK with the
vibrational decay rate $\Gamma_k/2\pi=1$ Hz. Other parameters employed are $\lambda/2\pi=0.115 $ MHz,
$\Gamma/2\pi=15$ MHz and $\Delta=0$.} \label{phase}
\end{figure}

Both the analytical and numerical results imply that our cooling
scheme is more powerful than previously proposed ones
\cite{epl-95-40003,prl-103-227203,
pra-85-033835,pra-86-053828,oe-21-029695}, particularly for the lower vibrational
frequency ($\omega_{k}/2\pi < 1$ MHz) and under the weaker laser field ($\Omega /2\pi < 3$ MHz).
Considering the numerical results in figures \ref{steady} and \ref{phase}, we observe that
the maximal cooling rate $W_{max}$ in our case can be more than $0.5\lambda$ once $\omega_{k}/2\pi\ge$2 MHz,
implying a much better cooling than in the resolved sideband regime in the laboratory representation \cite{prb-79-041302}. Moreover,
compared with the EIT cooling scheme \cite{oe-21-029695}, our scheme reaches the maximal cooling rate $W_{max}>0.5\lambda$
by a weaker cooling laser, e.g., approximately $\Omega /2\pi=2$ MHz, for the lower frequency vibrational
mode, e.g., $\omega _{k}/2\pi=2$ MHz. In contrast, reaching such a cooling rate $W$ in the EIT cooling scheme \cite{oe-21-029695}
requires $\omega _{k}/2\pi =10$ MHz and $\Omega /2\pi =80$ MHz, where the Rabi frequency $\Omega$ is
linearly proportional to the vibrational frequency $\omega_{k}$. So our scheme can release the demanding
experimental requirement of strong laser beams compared with the previous scheme. In addition, the
figures also present that cooling
the cantilever with $\omega_{k}/2\pi < 1$ MHz is less efficient compared with $\omega _{k}/2\pi=2$ or 3 MHz,
but still working and reaching $W_{max}\simeq 0.1816\lambda$, which guarantees that our method can be
faster to realize the ground state cooling of the cantilever then the one based on the microwave fields \cite{prb-79-041302} with the identical parameters.

\section*{Discussion}

\begin{figure}[tbph]
\centering\includegraphics[width=14cm,height=5cm]{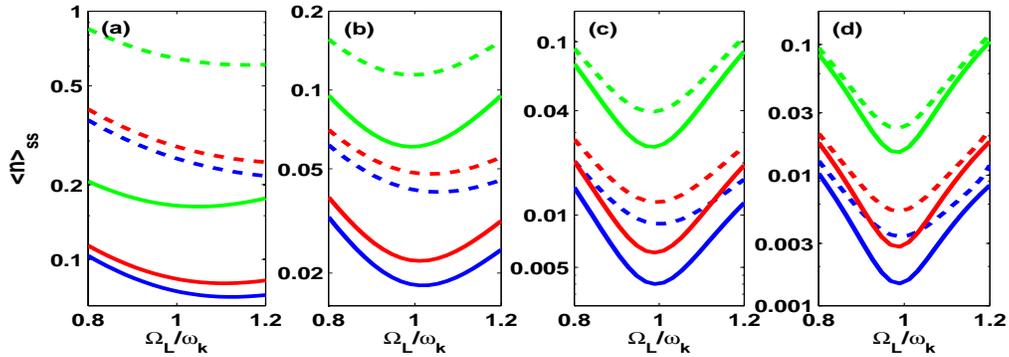}
\centering \caption{The final average phonon number $\langle
n\rangle_{ss}$ (in logarithmic scale) versus $ \Omega_L/\protect\omega_k$ for different
$\omega_k$, where (a) $\omega_k/2\pi=0.5$ MHz; (b) $\omega_k/2\pi=1$ MHz; (c) $\omega_k/2\pi=2$ MHz;
(d) $\omega_k/2\pi=3$ MHz. In every panel, the solid curves are simulated by
equation (\protect\ref{C11-2}) and the dash curves are the
numerical results by the master equation of $H^{rot}$. The pairs of
the curves, from the bottom to top, correspond to $\Gamma_k/2
\protect\pi=0$, 1 and 10 Hz, respectively. Other parameters take
the values as $\Omega/2\protect\pi=2$ MHz,
$\Gamma/2\protect\pi =15$ MHz, $T=20$ mK,
$\protect\lambda/2\protect\pi =0.115$ MHz and $\Delta=0$.}
\label{omegaton}
\end{figure}

To check how well our scheme works in a realistic system, we
consider below the variation of the cooling effect with respect to
the deviation from the work point $\Omega_{L}=\omega_{k}$ of the
Stark-shift gate. Figure \ref {omegaton} presents that the minimal
average phonon number $\langle n\rangle_{ss}$ changes slightly with
$\omega_k$, but this change becomes less evident for the larger $\omega_k$.
Since the work point only maximizes $A_-$, the fact that $\langle n\rangle_{ss}$
is determined by both $A_-$ and $A_+$ leads to the onset of the minimal average
phonon number deviated from the work point. For a given decay rate $\Gamma_{k}$,
the larger $\omega_k$ is more beneficial for cooling at the work point, and less
sensitive to the deviation from the work point. In summary, our proposed cooling
is robust against the experimental imperfection.

We should also assess the influence from the nuclear spin bath in
the NV center, which might seriously affect the final average phonon
number $\langle n\rangle_{ss}$ and the cooling time $t$. To this
end, we have considered some concrete values of the parameters, such
as $\langle n\rangle _{initial}=5,\omega _{k}/2\pi =2$ MHz, $\lambda
/2\pi =0.115$ MHz, $\Gamma /2\pi =15$ MHz, $T=20$ mK, $\Omega /2\pi
=2$ MHz, $\Omega _{L}=\omega _{k}$, $\Delta =0$ and $\Gamma
_{k}/2\pi =10$ Hz. Provided the nuclear spin bath taking the random
energy $\delta_{n}/2\pi\leq 0.1$ MHz ($0.5$ MHz), the final average
phonon number increases from $\langle n\rangle_{ss}=0.0399$ to
$0.0436$ ($0.1729$) and the corresponding cooling takes time from
$t=$ 39.7 $\mu$s to 42.2 $\mu$s (97.5 $\mu$s). Therefore,
suppressing the influence from the nuclear spin bath is very
important in order to achieve our cooling scheme. Possible
approaches include the dynamic nuclear polarization technology
\cite{prl-102-057403} and the isotopic purification of NV center
\cite{Nl-12-2083}, which have been widely adopted in the field of
spintronics.

In summary, we have studied an efficient cooling of the cantilever vibration by
a dynamic Stark-shift gate, in which the carrier transition and the
blue-sideband transition can be effectively suppressed when the
operation is made around the work point of the Stark-shift gate.
We have shown the possibility to cool the low-frequency cantilever down
to the vicinity of the ground state using weak cooling lasers.

Compared to the previous proposal for the trapped ion
\cite{njp-9-279}, our scheme has major differences in following two
points. The essential difference is the much larger mass of the
cantilever compared to the trapped ion. As a result, in our case the coupling of
the motional degrees of freedom of the cantilever to the spin
degrees of freedom of the NV center is provided by a strong magnetic
field gradient. In contrast, a laser radiation simply does this with an optical mechanical effect \cite{njp-9-279}. 
The second difference lies in different technical requirements.
For example, the effective coupling strength $\Omega_{L}$ is achieved by using a
Raman transition in our case. In a word, the dynamical Stark-shift in our case is from
the classical-field-assisted MFG, rather than the laser field for trapped ion \cite{njp-9-279}.

The analytical results we obtained, although not always accurate, have presented a clear
relationship of the final average phonon number with the vibrational
decay rate and the bath temperature. Moreover, we have also analyzed
the limitation of the maximal cooling rate. We argue that our scheme is experimentally
timely and relevant, and
would be useful for achieving an efficient cooling of the cantilever
vibration using currently available techniques.

\section*{Methods}

\textbf{The cooling mechanics based on the Stark-shift gate.} The cooling in our scheme is based on the Stark-shift gate. According to earlier works\cite{njp-9-279,pra-62-042307,prl-87-127901}, the Stark-shift gate in
the total Hamiltonian $H_{0}+V$ is described by
\begin{equation}
H_{ss}=\omega _{k}a^{\dagger }a+\frac{1}{2}\Omega _{L}(\left\vert
b\right\rangle \left\langle b\right\vert -\left\vert d\right\rangle
\left\langle d\right\vert )+V.  \label{B1}
\end{equation}
In the interaction picture, after the rotating wave approximation is
applied, the Hamiltonian in (\ref{B1}) can be rewritten as
\begin{equation}
H_{I}=\lambda (\left\vert b\right\rangle \left\langle d\right\vert
ae^{i\delta t}+\left\vert d\right\rangle \left\langle b\right\vert
a^{\dagger }e^{-i\delta t}),  \label{B2}
\end{equation}
with $\delta =\Omega _{L}-\omega _{k}$. When $\Omega _{L}=\omega _{k}$,
i.e., the work point of the Stark-shift gate,
the above Hamiltonain reduces to
\begin{equation}
H_{I}=\lambda (\left\vert b\right\rangle \left\langle d\right\vert
a+\left\vert d\right\rangle \left\langle b\right\vert a^{\dagger}),
\label{B21}
\end{equation}
which is a typical Jaynes-Cummings interaction for
the first-order red-sideband transition between the dark and bright
states, leading to the phonon number change in the
cantilever vibration.

\begin{figure}[tbp]
\centering
\includegraphics[width=12cm,height=5.5cm]{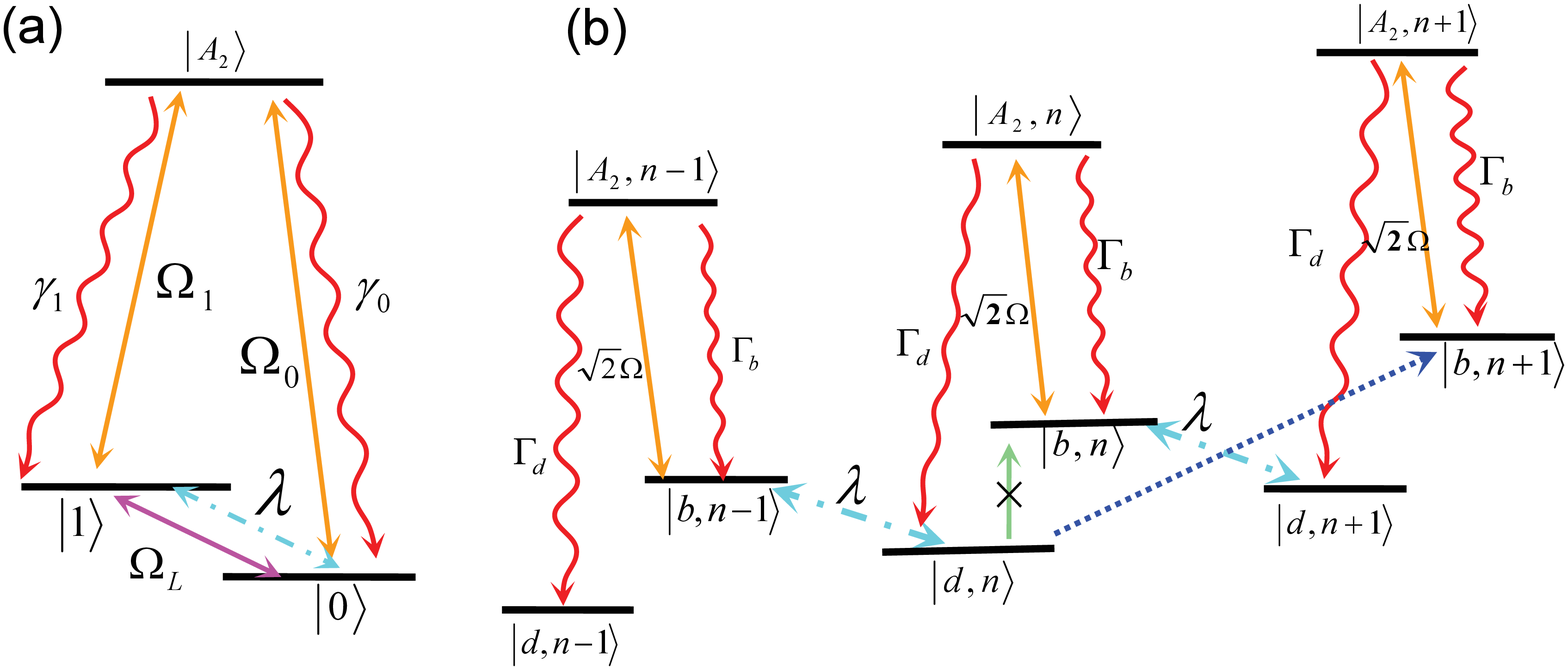}
\caption{(a) The effective cooling cycle in our scheme. (b) The schematic
for the cooling mechanism. The Stark-shift-gate assisted cooling includes an effective
classical field creating the Stark shift of the ground state, a strong MFG
whose coupling $\protect\lambda $ leads to the red-sideband transition $%
\left\vert d,n\right\rangle \leftrightarrow \left\vert b,n-1\right\rangle $,
and the optical lasers coupling the bright state $|b\rangle $ to the excited
state $|A_{2}\rangle$. The optical lasers should satisfy the two-phonon
resonance condition, which yields elimination of the carrier transition
between $|d,n\rangle$ and $|b,n\rangle$ and suppression of the blue-sideband
transition between $|d,n\rangle$ and $|b,n+1\rangle $. So the strong MFG
only contributes to the red-sideband transition. In our treatment, we
suppose $\Gamma_{d}=\Gamma_{b}$ and $\Gamma=\protect\gamma_{1}+\protect\gamma%
_{2}$, and other parameters are defined in the text or in figure 1.}
\label{cool}
\end{figure}

The cooling process in our scheme is described in figure \ref{cool}. In
terms of equation (\ref{B21}), assume the system initially in the state $\left\vert
d,n+1\right\rangle$, the only possible transition is from $\left\vert
d,n+1\right\rangle$ to $\left\vert b,n\right\rangle $, which is caused by the
Stark-shift gate. It also implies that the blue-sideband transition and the
carrier transition relevant to the dark state $\left\vert d\right\rangle$
are suppressed due to the Stark-shift gate. In the laboratory frame, the transition $\left\vert
b,n\right\rangle \leftrightarrow\left\vert A_{2},n\right\rangle$ is actually driven
by two lasers with the same Rabi frequencies and detunings. In the rotating frame
as plotted in figure \ref{cool}, the spin state
is first excited to $\left\vert A_{2},n\right\rangle$, and then decays to
the bright state $|b,n\rangle $ or the dark state $\left\vert
d,n\right\rangle$. If the decay is to the dark state $\left\vert
d,n\right\rangle$, a phonon is lost from the cantilever vibration due to $H_{I}$
and the cooling goes to the next step. But if the decay is to
the bright state $\left\vert b,n\right\rangle$, the state will be pumped to
the excited state $\left\vert A_{2},n\right\rangle$ again, and this cycle of
the laser cooling will repeat until the decay is to the dark state.

A clearer picture for above cooling process with the phonon dissipation
governed by the transition $\left\vert
d,n+1\right\rangle\leftrightarrow\left\vert b,n\right\rangle$ can be found
in Supplementary Information by numerical calculations of the absorption spectra. We may
find that the carrier transition $\left\vert d,n\right\rangle \rightarrow
\left\vert b,n\right\rangle $ is totally suppressed by the destructive
interference, and the blue-sideband transition $\left\vert
d,n\right\rangle\rightarrow \left\vert b,n+1\right\rangle $ is largely
suppressed. As a result, repeating the laser cooling cycles, the cantilever
will finally be cooled down to the vibration ground state.

Before going further to the numerical calculation, we simply compare the Stark-shift-gate cooling with the EIT
cooling\cite{oe-21-029695}. From the schematic illustrations, both of them share
the similar quantum interference process and steady state, which can effectively
suppress the  blue-sideband transition and the carrier transition. Besides, the Stark-shift-gate cooling
goes beyond the EIT cooling by an additional coupling, which drives the transition between
the dark and bright states, constituting a Jaynes-Cummings interaction for
the first-order red-sideband transition at the work point $\Omega _{L}=\omega _{k}$. As a result,
different from the EIT cooling, the Stark-shift-gate cooling works with the efficiency independent
from the cooling laser strength, but mainly relevant to the work point.


\begin{thebibliography}{99}

\bibitem{Sicience-330-1520} Weis, S. et al. Optomechanically Induced Transparency. \textit{Science} \textbf{330}, 1520-1523 (2010).

\bibitem{nature-436-87} Birnbaum, K. M. et al. Photon blockade in an optical cavity with one trapped atom. \textit{Nature} \textbf{436}, 87-90 (2005).

\bibitem{prl-107-063601} Rabl, P. Photon Blockade Effect in Optomechanical Systems. \textit{Phys. Rev. Lett.} \textbf{107}, 063601 (2011).

\bibitem{pra-32-2287} Imoto, N., Haus, H. A. \& Yamamoto, Y. Quantum nondemolition measurement of the photon number via the optical Kerr effect. \textit{Phys. Rev. A} \textbf{32}, 2287 (1985).

\bibitem{V.B.Braginsky} Braginsky, V. B. \& Manukin, A. B. \textit{%
Measurements of Weak Forces in Physics Experiments} Edited by Douglass D H,
(Chicago University Press, Chicago, 1977).

\bibitem{prb-88-085201} Chotorlishvili, L. et al. Entanglement between nitrogen vacancy spins in diamond controlled by a nanomechanical resonator. \textit{Phys. Rev. B} \textbf{88}, 085201 (2013).

\bibitem{prb-68-155311} Irish, E. K. \& Schwab, K. Quantum measurement of a coupled nanomechanical resonator–Cooper-pair box system. \textit{Phys. Rev. B} \textbf{68}, 155311 (2003).

\bibitem{nature-459-960} LaHaye, M. D., Suh, J., Echternach, P. M., Schwab, K. C. \&
Roukes, M. L. Nanomechanical measurements of a superconducting qubit. \textit{Nature} \textbf{459}, 960-964 (2009).

\bibitem{natureN-3-691} Ndieyira, J. W. et al. Nanomechanical detection of antibiotic–mucopeptide binding in a model for superbug drug resistance. \textit{Nat. Nanotech.}
\textbf{11} 691 - 696 (2008).

\bibitem{natN-3-501} Tetard, L et al. Imaging nanoparticles in cells by nanomechanical holography . \textit{Nat. Nanotech.} \textbf{8}, 501 (2008).

\bibitem{prl-99-140403} Treutlein, P. et al. Bose-Einstein Condensate Coupled to a Nanomechanical Resonator on an Atom Chip. \textit{Phys. Rev. Lett.} \textbf{99}, \ 140403 ( 2007).

\bibitem{prl-92-075507} Wilson-Rae, I., Zoller, P. \& Imamoglu, A.  Laser Cooling of a Nanomechanical Resonator Mode to its Quantum Ground State. \textit{Phys. Rev. Lett.} \textbf{92}, \ 075507 (2004).

\bibitem{nature-475-359} Teufel, J. D. et al. Sideband cooling of micromechanical motion to the quantum ground state. \textit{Nature} \textbf{475}, 359-363 (2011).

\bibitem{prb-78-134301} Li, Y., Wang, Y. D., Xue, F. \& Bruder, C. Cooling a micromechanical beam by coupling it to a transmission line. \textit{Phys. Rev. B} \textbf{78}, 134301 (2008).

\bibitem{nature-432-200} Yoshie, T. et al. Vacuum Rabi splitting with a single quantum dot in a photonic crystal nanocavity.  \textit{Nature}
\textbf{432}, 200-203 (2004).

\bibitem{prl-99-093901} Wilson-Rae, I., Nooshi, N., Zwerger, W. \& Kippenberg, T.
J. Theory of Ground State Cooling of a Mechanical Oscillator Using Dynamical Backaction. \textit{Phys. Rev. Lett.} \textbf{99}, 093901 (2007).

\bibitem{prl-99-093902} Marquardt, F., Chen, J. P., Clerk, A. A. \& Girvin, S. M. Quantum Theory of Cavity-Assisted Sideband Cooling of Mechanical Motion. \textit{Phys. Rev. Lett.} \textbf{99}, 093902 (2007).

\bibitem{prb-76-205302} Xue, F., Wang, Y. D., Liu, Y. X. \& Nori, F. Cooling a micromechanical beam by coupling it to a transmission line. \textit{%
Phys. Rev. B} \textbf{76}, \ 205302 (2007).

\bibitem{nature-443-193} Naik, A. et al. Cooling a nanomechanical resonator with quantum back-action. \textit{Nature} \textbf{443}, 193-196 (2006).

\bibitem{prl-108-120602} Mari, A. \& Eisert, J. Cooling by Heating: Very Hot Thermal Light Can Significantly Cool Quantum Systems. \textit{Phys. Rev. Lett.}
\textbf{108}, 120602 (2012).

\bibitem{JPCM-25-142201} Zhang, J. Q., Li, Y. \& Feng, M. Cooling a charged mechanical resonator with time-dependent bias gate voltages. \textit{
J.Phys.:Condens. Matter} \textbf{25}, 142201 (2013).

\bibitem{pra-83-043804} Li, Y., Wu, L. A. \& Wang, Z. D. Fast ground-state cooling of mechanical resonators with time-dependent optical cavities. \textit{Phys. Rev. A}
\textbf{83}, 043804 (2011).

\bibitem{pra-85-025804} Deng, Z. J., Li, Y., Gao, M.  \& Wu, C. W. Performance of a cooling method by quadratic coupling at high temperatures. \textit{Phys.
Rev. A} \textbf{85}, 025804 (2012).

\bibitem{prb-84-094502} Li, Y, Wu, L. A., Wang, Y. D. \& Yang, L. P. Nondeterministic ultrafast ground-state cooling of a mechanical resonator. \textit{%
Phys. Rev. B} \textbf{84}, 094502 (2011).

\bibitem{epl-95-40003} Li, Z. Z., Ouyang, S. H., Lam, C. H. \& You, J. Q. Cooling a nanomechanical resonator by a triple quantum dot.  \textit{%
Europhys. Lett.} \textbf{95}, 40003 (2011).

\bibitem{prl-103-227203} Xia, K. Y. \& Evers, J. Ground State Cooling of a Nanomechanical Resonator in the Nonresolved Regime via Quantum Interference. \textit{Phys. Rev. Lett.}
\textbf{103} 227203 (2009).

\bibitem{pra-85-033835} Zhu, J. P., Li, G. X. \& Ficek, Z. Two-particle dark-state cooling of a nanomechanical resonator. \textit{Phys. Rev.
A} \textbf{85} 033835 (2012).

\bibitem{pra-86-053828} Zhu, J. P. \& Li, G. X. Ground-state cooling of a nanomechanical.  \textit{Phys. Rev. A}
\textbf{86} 053828 (2012).

\bibitem{oe-21-029695} Zhang, J. Q. et al. Fast optical cooling of nanomechanical
cantilever with the dynamical Zeeman
effect \textit{Optics Express} \textbf{21}, 029695 (2013).

\bibitem{prl-85-4458} Morigi, G., Eschner, J. \& Keitel C. H. Ground State Laser Cooling Using Electromagnetically Induced Transparency. \textit{Phys. Rev. Lett.} \textbf{85} 4458 (2000).

\bibitem{prl-85-5547} Roos, C. F. et al. Experimental Demonstration of Ground State Laser Cooling with Electromagnetically Induced Transparency.  \textit{Phys. Rev. Lett.} \textbf{85}, 5547 (2000).

\bibitem{njp-9-279} Retzker, A. \& Plenio, M. B. Fast cooling of trapped ions using the dynamical
Stark shift. \textit{New J. Phys.}
\textbf{9}, 279 (2007).

\bibitem{prb-79-041302} Rabl, P. et al. Strong magnetic coupling between an electronic spin qubit and a mechanical resonator. \textit{Phys. Rev. B} \textbf{79}, 041302 (2009).

\bibitem{natphys-7-879} Arcizet, O. et al. A single nitrogen-vacancy defect coupled to a nanomechanical oscillator. \textit{Nat. Phys.} \textbf{11}, 879-883 (2011).

\bibitem{njp-13-025025} Maze, J. R. et al. Properties of nitrogen-vacancy centers in diamond: the group theoretic approach. \textit{New J. Phys.} \textbf{13}, 025025 (2011).

\bibitem{nature-466-730} Togan, E. et al. Quantum entanglement between an optical photon and a solid-state spin qubit.
\textit{Nature} \textbf{466}, 730-734 (2010).

\bibitem{pra-83-054306} Chen, Q., Yang, W. L., Feng, M. \& Du, J. F. Entangling separate nitrogen-vacancy centers in a scalable fashion via coupling to microtoroidal resonators. \textit{%
Phys. Rev. A} \textbf{83}, 054305 (2011).

\bibitem{prl-111-227602} MacQuarrie, E. R., Gosavi, T. A., Jungwirth, N. R., Bhave, S. A. \& Fuchs, G. D. Mechanical Spin Control of Nitrogen-Vacancy Centers in Diamond. \textit{Phys. Rev. Lett.} \textbf{111}, 227602 (2013).

\bibitem{jpb-36-1041} Morigi, G. \& Eschner, J. Is an ion string laser-cooled like a single ion? \textit{J. Phys. B: At.
Mol. Opt. Phys.} \textbf{36}, 1041 (2003).

\bibitem{prl-87-257904} Mintert, F. \& Wunderlich, C. Ion-Trap Quantum Logic Using Long-Wavelength Radiation. \textit{Phys. Rev. Lett. } \textbf{87}, 257904 (2001).

\bibitem{Gardiner} Gardiner, S. A.  \textit{Quantum Measurement, Quantum
Chaos, and Bose-Einstein Condensates} Dissertation (Leopold-Franzens-Universitat Innsbruck, 1977).

\bibitem{pra-62-042307} Jonathan, D., Plenio, M. B. \& Knight, P. L. Fast quantum gates for cold trapped ions. \textit{%
Phys. Rev. A} \textbf{62}, 042307 (2000).

\bibitem{prl-87-127901} Jonathan, D. \& Plenio, M. B. Light-Shift-Induced Quantum Gates for Ions in Thermal Motion. \textit{Phys. Rev. Lett} \textbf{87}, 127901 (2001).


\bibitem{pra-67-033402} Morigi, G. Cooling atomic motion with quantum interference. \textit{Phys. Rev. A} \textbf{67},
033402 (2003).

\bibitem{pra-72-043823} Rabl, P., Steixner, V. \& Zoller, P. Quantum-limited velocity readout and quantum feedback cooling of a trapped ion via electromagnetically induced transparency. \textit{Phys.
Rev. A.} \textbf{72}, 043823 (2005).

\bibitem{pra-49-2771} Marzoli, I., Cirac, J. I., Blatt, R. \& Zoller, P. Laser cooling of trapped three-level ions: Designing two-level systems for sideband cooling. \textit{Phys.
Rev. A.} \textbf{49}, 2771 (1994).

\bibitem{pra-85-032111} Reiter, F. \& Sorensen, A. S. Effective operator formalism for open quantum systems. \textit{Phys.
Rev. A.} \textbf{85}, 032111 (2012).

\bibitem{prl-102-057403} Jacques, V. et al. Dynamic Polarization of Single Nuclear Spins by Optical Pumping of Nitrogen-Vacancy Color Centers in Diamond at Room Temperature.
\textit{Phys. Rev. Lett.} \textbf{102}, 057403 (2009).

\bibitem{Nl-12-2083} Ishikawa, T. et al. Optical and Spin Coherence Properties of Nitrogen-Vacancy Centers Placed in a 100 nm Thick Isotopically Purified Diamond Layer. \textit{Nano Lett.} \textbf{12},
2083-87 (2012).

\end{thebibliography}

\section*{Acknowledgements}

JQZ thanks Xing Xiao, and Nan Zhao for helpful discussions. This work is supported by
National Fundamental Research Program of
China (Grants No. 2012CB922102 and No. 2013CB921803), National
Natural Science Foundation of China (Grants Nos. 11274352, 91421111 and
11304366) and the China Postdoctoral Science Foundation (Grants
Nos. 2013M531771 and 2014T70760).

\section*{Author contributions statement}
LLY contributed to numerical and prepared the first version of the
manuscript,  JQZ and SZ designed this work, JQZ and MF gave physical analysis,
LLY, JQZ and MF wrote the manuscript.

\section*{Additional information}
The authors declare no competing financial interests. Supplementary information accompanies this paper is enclosed in another pdf file.

\end{document}